\documentclass[twocolumn,showpacs,aps]{revtex4}
\usepackage{graphicx}
\usepackage{psfrag}
\usepackage{dcolumn}
\usepackage{bm}
\usepackage{amsmath}
\usepackage{amsfonts}
\usepackage{amssymb}
\usepackage[latin1]{inputenc}

\newcommand{\ket}[1]{|#1\rangle}
\newcommand{\pket}[1]{\Vert #1\rangle}
\newcommand{\bra}[1]{\langle#1|}

\newcommand{\op}[1]{|#1\rangle\langle#1|}
\newcommand{\opp}[2]{|#1\rangle\langle#2|}
\newcommand{\popp}[2]{\Vert #1\rangle\langle#2\Vert}

\newcommand{\mc}[1]{{\mathcal #1}}

\begin{document}

\title{Quantum Walk and Iterated Quantum Games}

\author{G. Abal, R. Donangelo}\thanks{Permanent address:
Instituto de Física, Universidade Federal do Rio de Janeiro, C.P. 68528, Rio de Janeiro, 21941-972, Brazil.}\author{H. Fort}
\affiliation{Instituto de Física, Universidad de la República \\
C.C. 30, C.P. 11300, Montevideo, Uruguay }

\date{\today}

\begin{abstract}
Iterated bipartite quantum games are implemented in terms of the discrete-time quantum walk on the line. Our proposal allows for conditional strategies, as two rational agents make a choice from a restricted set of two qubit unitary operations. Several frequently used classical strategies give rise to families of corresponding quantum strategies. A quantum version of the Prisoner's Dilemma in which both players use mixed strategies is presented as a specific example. Since there are now quantum walk physical implementations at a proof-of principle stage, this connection may represent a step towards the experimental realization of quantum games.
\end{abstract}



\maketitle

\section{Introduction}

Quantum Walks (QWs) \cite{Kempe-review} have captured the attention of
quantum information theorists mainly because of their potential for the development of new quantum algorithms \cite{Kempe02,Shenvi03,Childs04,Fahri}. When two independent quantum walks are considered, their joint state
may be entangled in several ways, and this opens interesting possibilities for quantum information processing. Quantum
walks have been realized using technologies ranging from  NMR processors \cite{Ryan} to low-intensity linear optics experiments \cite{Do}. Quantum Games (QGs) \cite{Meyer99,Eisert99,Iqbal02,lj03} constitute another approach to quantum information processing where quantum players can achieve results not available to classical players \cite{Eisert99,Du03}. In particular, QGs may provide a new persective on the long-standing ``public goods'' distribution problem \cite{Patel06,ChenHogg02}. A recent experiment,  involving trained rational human subjects, indicates that different cooperation levels (with respect to the corresponding classical game) are to be expected in a quantum version of the Prisioner's Dilemma, both in the one-shot case and in an iterated version \cite{ChenHogg06}.

In this work, the QWs and QGs approaches to quantum information processing are related. This is done by using the QW to formulate iterated quantum games in which conditional strategies ({\it i.e.} strategies that depend on the previous state of {\it both} players) are  naturally implemented. In section II we briefly introduce the discrete-time quantum walk  and provide some background material about classical, bi-partite, non-zero sum games. In Section III we introduce a simple set of rules which allows the construction of a quantum game based on two quatum walkers. We also discuss the possible strategies and present some examples. In Section IV we present our conclusions. 

\section{Preliminary concepts}
Before establishing a connection between QWs and QGs, it is necessary to establish a few definitions and provide some reference material on both systems. 

\subsection{Discrete-time quantum walk}

The Hilbert space of a quantum walk on a line is composed of two parts, $\mc{H}=\mc{H}_x\otimes \mc{H}_c$. The one-qubit ``coin" subspace, $\mc{H}_c$, is spanned by two orthonormal states $\{\ket{0}, \ket{1}\}$.  The spatial subspace, $\mc{H}_x$, is spanned by the orthonormal set of position eigenstates, $\{\pket{ x}\}$, with $x\in\mathcal Z$ labelling discrete sites on a line. We use the symbol $\pket{\cdot}$ to distinguish the position eigenstates from the kets in $\mc{H}_c$. The evolution is generated by repeated application of a composite unitary operator $U$ which implements a coin operation, followed by a conditional shift in the position of the walker. This shift operation entangles the coin and position of the walker \cite{Abal06}. The quantum walk with two walkers $A,B$ \cite{2dwalks} takes place in a Hilbert space $\mc{H}_{AB}=\mc{H}_A\otimes \mc{H}_B$, where $\mc{H}_A$ and $\mc{H}_B$ are isomorphous to $\mc{H}$.  After $N$ steps, a pure state characterized by a density operator $\rho_0=\op{\Psi(0)}$ evolves to $\rho_N=U^N\rho_0\left(U^\dagger\right)^N$ with
\begin{equation}
U=\Omega\cdot\left(I\otimes U_c\right), \label{evol}
\end{equation}
where $U_c$ is a unitary operation in $\mc{H}_c^{\otimes 2}$, $I$ is the identity in $\mc{H}_x^{\otimes 2}$ and
$\Omega$ is a shift operation in $\mc{H}_{AB}$. We shall be concerned with non-separable coin operations $U_c$, {\it
i.e.} which cannot be written as products of operations local to subspaces $\mc{H}_A, \mc{H}_B$. A general conditional shift operation may be expressed in terms of position eigenstates
$\pket{x_A,x_B}\equiv\pket{x_A}\otimes\pket{x_B}$ as
\begin{eqnarray} 
\Omega&\equiv&\sum_{x_A,x_B}\left\{\popp{x_A+s_A^{(0)},x_B+s_B^{(0)}}{x_A,x_B}
\otimes\opp{00}{00}+\right.\nonumber\\
&&\quad\popp{x_A+s_A^{(1)},x_B+s_B^{(1)}}{x_A,x_B}
\otimes\opp{01}{01}+\nonumber\\
&&\quad\popp{x_A+s_A^{(2)},x_B+s_B^{(2)}}{x_A,x_B}
\otimes\opp{10}{10}+\nonumber\\
&&\quad\left.\popp{x_A+s_A^{(3)},x_B+s_B^{(3)}}{x_A,x_B}
\otimes\opp{11}{11}\right\}.\label{shift-op-gen}
\end{eqnarray}
The sums are over all discrete sites $(x_A,x_B)$ on a plane and the integer parameters $s_{A,B}^{(i)}\;  (i=0\ldots 3)$
denote the step sizes taken by each walker for each coin state. Note that this shift operator is non-separable with
respect to subspaces $\mc{H}_A$ and $\mc{H}_B$, except in the particular case in which the parameters satisfy the relations 
\begin{equation}
s_A^{(0)}=s_A^{(1)},\quad s_A^{(2)}=s_A^{(3)},\quad s_B^{(0)}=s_B^{(2)},\quad s_B^{(1)}=s_B^{(3)}.  
\end{equation} 

The QW model defined by eqs.~(\ref{evol}) and (\ref{shift-op-gen}) is our starting
point for establishing a connection with iterated quantum games. Before addressing this issue, we summarize some relevant information on non-zero sum classical games.\\

\subsection{Iterated games and conditional strategies} 

In classical, bi-partite,  non-zero sum games a gain by one agent does not imply a loss by the other. The paradigmatic example of this situation is the Prisoner's Dilemma (PD) \cite{f52} in which each agent is confronted with the option to ``cooperate" (C) or ``defect" (D). Both players obtain a ``reward" $R$ if they both play $C$, but if one of them deviates and plays $D$, he obtains the temptation $T>R$. If both players defect, they are penalized and lose $P$, but the worst payoff is obtained by a cheated cooperator who gets $S$.  In sum, the PD payoffs obey
\begin{equation}
T>R>P>S \quad\mbox{ and }\quad 2R>T+S\label{ineq}
\end{equation}
so that, on the average, it pays more to defect $(T+P>R+S)$ and [D,D] is a Nash equilibrium (NE), {\it i.e.} in which none of the two players can improve his/her payoff through a unilateral change in strategy. On the other hand, in the case of [C,C] no player can improve his/her payoff without worsening the other player's payoff, so it is Pareto optimal (PO). The dilemma consists in the fact that the NE is not PO.

When the PD game is played repeatedly, each player has the opportunity to retaliate (reward) the other for having
played D (C) in the previous encounter. Iterated games introduce the possibility of developing conditional
strategies, and mutual cooperation may arise as an equilibrium outcome. In this context, a classical strategy is usually
characterized by four conditional probabilities $[p_R,p_S,p_T,p_P]$, where $p_i$ is the probability of choosing C after having received a payoff $i=R, S, T\mbox{ or }P$, respectively. For instance, the {\it win-stay, lose-shift} or {\it Pavlov} strategy \cite{kk88}, represented by $[1,0,0,1]$, corresponds to flipping the state when the previous payoff is unsatisfactory  ($P$ or $S$). Similarly, an unbiased random strategy is described by
$[\frac12,\frac12,\frac12,\frac12]$ and the TFT (tit-for-tat)  strategy, in which a player copies the previous move of the other player, is represented by $[1,0,1,0]$.\\

\section{Quantum walk as a quantum game} 

The essential observation is that the operator $\Omega$, defined in
eq. (\ref{shift-op-gen}), connects particular coin states with the corresponding payoffs $s_{A,B}$. When the position variables $x_{A,B}$ are associated to the accumulated payoffs, an iterated quantum game may be constructed where each player applies a certain strategy acting on the coin subspace, as defined below. The number of games that may be defined in this way is huge. For illustrative purposes, in what follows we use this prescription to construct a quantum version of the iterated PD game. However, it is important to emphasize that the connection between
QWs and QGs is rather general and other classical games may be constructed and analyzed along similar lines.

Consider two agents A (Alice) and B (Bob) which act as opponents in an iterated quantum game based on the PD. Quantum versions of iterated PD games are obtained from the QW by three simple rules:
\begin{enumerate}
\item[i.] The coin states of each agent are interpreted as $\protect{\ket{0}\rightarrow C}$ (cooperation) and $\protect{\ket{1}\rightarrow D}$ (defection).
\item[ii.] Each agent is allowed to act on his/her own ``coin" qubit with a unitary operation
(a strategy) $U_A$ or $U_B$ in $\mc{H}^{\otimes 2}$. However, an agent is not allowed to alter his/her opponent's state with this operation.
\item[iii.] The position subspace is a quantum register in which the accumulated payoff of each agent is stored. If $X_A$ is the position operator for Alice, $\protect{X_A\pket{x_A}=x_A\pket{x_A}}$, her average payoff is $\protect{\bar x_A =\mbox{tr}(\rho X_A)}$, and the same holds for Bob.
\end{enumerate}
The position registers are updated after each coin operation, according to eq.~(\ref{shift-op-gen}) with the following identifications
\begin{eqnarray}
s_A^{(0)}=s_B^{(0)}=R,&\quad& s_A^{(1)}=s_B^{(2)}=S,\nonumber\\
 s_A^{(2)}=s_B^{(1)}=T,&\quad &s_A^{(3)}=s_B^{(3)}=P.
\end{eqnarray} 
These relations, according to condition (\ref{ineq}), imply that the shift operator $\Omega$ for the quantum iterated PD game is non-separable with respect to subspaces $A,B$. We assume that both players start in the position eigenstate, $\protect{\ket{x_A,x_B}=\ket{0,0}}$ with arbitrary initial coin $\ket{c_0}\in \mc{H}_c^{\otimes 2}$. A
pure initial state $\protect{\rho_0=\popp{0,0}{0,0}\otimes\op{c_0}}$ evolves to $\rho_N$ after $N$ applications of $U$, defined in eq.~(\ref{evol}). Then, a measurement of the observables $X_{A,B}$ determines the final payoff for each
player. Alternatively, the average payoff $\bar x_{A,B}=\langle X_{A,B}\rangle$ after $N$ iterations can be used to
determine who did best. As expected, when partial measurements of the joint coin state are performed before each shift
operation, the average payoffs of the corresponding classical PD game are recovered. At the heart of the quantum game
is the choice of strategy made by each agent, within the restrictions of rule (ii) above. The operations $U_A$ or $U_B$ may
represent classical-like strategies,  but they may also account for new strategies with no classical analog. In this
context, it is of interest to explore the strategic choices available for both players and how do they relate to those
available in the classical iterated PD game.

\subsection{Sequential vs. simultaneous games}

Assuming that the first qubit from the left is Alice's and the second is Bob's, the possible strategies available to
Alice are represented by the set of unitary two-qubit operations that do not alter the second qubit,
\begin{eqnarray}
U_A&=&\left[a_0\ket{00}+a_1\ket{10}\right]\bra{00}+
 \left[a_2\ket{01}+a_3\ket{11}\right]\bra{01}+ \nonumber\\
&&\left[a_4\ket{00}+a_5\ket{10}\right]\bra{10}+ \left[a_6\ket{01}+a_7\ket{11}\right]\bra{11},\label{UAket}
\end{eqnarray}
where the complex coefficients $a_i$ satisfy the requirements for unitarity of $U_A$. Similarly, the possible strategies available to Bob are the set of unitary two-qubit operations that do not alter the first qubit
\begin{eqnarray}
U_B&=&\left[b_0\ket{00}+b_1\ket{01}\right]\bra{00}+
 \left[b_2\ket{00}+b_3\ket{01}\right]\bra{01}+ \nonumber\\
&&\left[b_4\ket{10}+b_5\ket{11}\right]\bra{10}+
\left[b_6\ket{10}+b_7\ket{11}\right]\bra{11}\label{UBket}
\end{eqnarray}
with the $b_i$ satisfying the requirements for the unitarity of $U_B$. These operations allow for conditional strategies, in which a player's action depends on the previous coin state of {\it both} players. Unconditional strategies result from separable coin operations, when $U_A$ and $U_B$ both reduce to local operations in $\mc{H}_c$.

In the general case, $[U_A,U_B]\ne 0$, and the order in which the operations are applied makes
a difference since the coin operation in eq.~(\ref{evol}) can be constructed either as $\protect{U_B\cdot U_A}$, if Alice moves first, or $\protect{U_A\cdot U_B}$ otherwise. We shall refer to these games, with composite coin operations applied in a prescribed order, as \emph{sequential games}. Another alternative is that both agents apply
their operations simultaneously. This gives rise to a \emph{simultaneous} version of the quantum game in which the coin operation $U_c$, a unitary operation in $\mc{H}_c^{\otimes 2}$, reflects the strategic choice of both players. A sequential game with separable coin operations is identical with the corresponding simultaneous game. However, for non-separable coin operations, some sequential games cannot be played simultaneously and, conversely, some simultanous
games cannot be played sequentially. We first discuss in some detail the strategies available in sequential games, as they are easier to visualize.

\begin{figure}
  \begin{center}
     \includegraphics[scale=0.4]{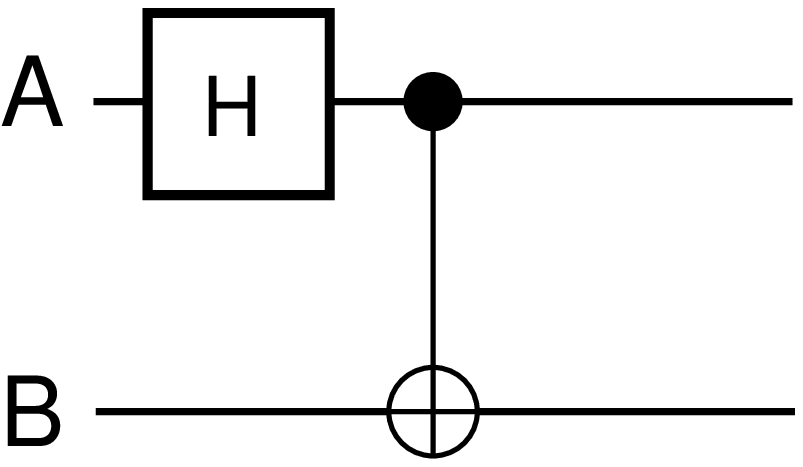}\hskip16mm\includegraphics[scale=0.4]{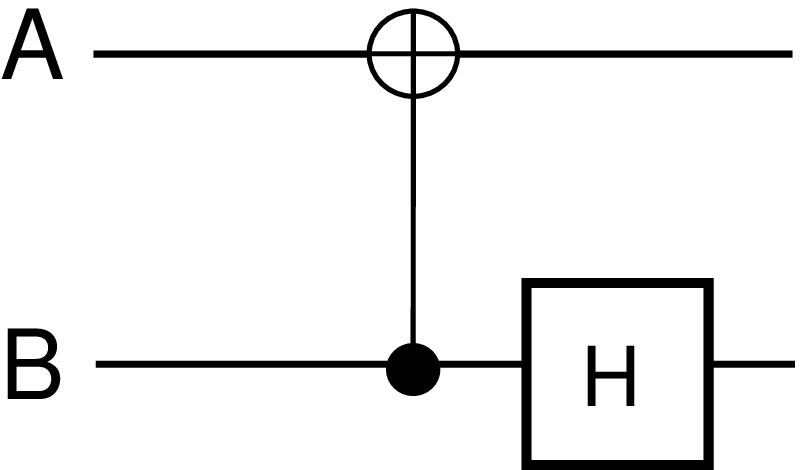}
   \end{center}
\caption{\footnotesize  Circuits representing the coin operation of a Pavlov vs. Random quantum game. Left: Alice plays Random and Bob responds with Pavlov. Right: Alice plays Pavlov and Bob plays Random. These circuits map computational basis states into maximally entangled (Bell) states. } \label{fig:cnot-bell}
\end{figure}

\subsection{Sequential games}

According to rule (ii) above, the quantum strategies available to Alice in a sequential game, are represented by unitary operations of the form ~(\ref{UAket}) affecting only the first qubit. The relation with a classical strategy, defined by four real parameters $[p_R,p_S,p_T,p_P]$ with $p_i\in[0,1]$, is made explicit with the parametrization
\begin{equation}
\begin{array}{lcl}
a_0=e^{i\varphi_R}\sqrt{p_R}&\qquad\qquad&a_1=e^{i\theta_R}\sqrt{1-p_R}\\
a_2=e^{i\varphi_S}\sqrt{p_S}&\qquad\qquad&a_3=e^{i\theta_S}\sqrt{1-p_S}\\
a_4=e^{i\varphi_T}\sqrt{p_T}&\qquad\qquad&a_5=e^{i\theta_T}\sqrt{1-p_T}\\
a_6=e^{i\varphi_P}\sqrt{p_P}&\qquad\qquad&a_7=e^{i\theta_P}\sqrt{1-p_P}
\end{array}\label{alice-set}
\end{equation}
where the phases $\varphi_i,\theta_i \in[-\pi,\pi]$. A similar parametrization holds for the coefficients $b_i$ in eq.~(\ref{UBket}).  Unitarity implies (aside from two conditions relating the phases) that
\begin{equation}\label{useq-strat}
p_R+p_T=p_S+p_P=1.
\end{equation}
%
%
Only classical strategies satisfying eq.~(\ref{useq-strat}) can be implemented as unitary operations in sequential quantum games. Each of these gives rise to a ``family" of related quantum strategies. Aside from a global phase, a given
sequential quantum strategy is determined by up to seven real parameters. Families of related strategies require fewer
parameters however. For instance, quantum versions of Pavlov's strategy as played by Alice, $[1,0,0,1]$, are
implemented with operators of the form
$$
U_A^P=\opp{00}{00}+e^{i\nu_1}\opp{11}{01}+e^{i\nu_2}
\opp{10}{10}+e^{i\nu_3}\opp{01}{11}
$$
in terms of three arbitrary phases $\nu_j$. If $\nu_j=0$, $U_A^{P}$ reduces to a controlled-NOT (CNOT) operation in
which Bob's coin is the control qubit \cite{N+C}.  Similarly, Pavlov's strategies played by Bob are of the form
$$
U_B^P=\opp{00}{00}+e^{i\mu_1}\opp{01}{01}+e^{i\mu_2} \opp{11}{10}+e^{i\mu_3}\opp{10}{11},\label{UBP}
$$
and for $\mu_j=0$, they reduce to a CNOT operation controlled by Alice's qubit. These are examples of conditional
strategies, based on non-separable operators. Furthermore, since $[U_A^{P},U_B^{P}]\ne 0$, the order makes a
difference when both players use Pavlov strategies.

ASan example of a separable strategy, consider the Hadamard operator $H$, defined by
$\protect{H\ket{k}=\frac{1}{\sqrt{2}}(\ket{0}+(-1)^k\ket{1})}$ for $k=0,1$, which has been widely used as a coin
operation in QW's \cite{Kempe-review}. In the context of QG's, the fact that it generates unbiased superpositions of
the computational basis states makes it useful as a quantum version of the classical {\it Random} strategy
$\left[\frac12,\frac12,\frac12,\frac12\right]$. A sequential quantum game in which Alice plays Pavlov and Bob replies
with Random, is described, for a particular choice of phases, by a coin operation $\protect{U_c=(I_1\otimes H)\cdot
U_A^{P}}$, represented by the circuit in the right panel of Fig.~\ref{fig:cnot-bell}.

\subsubsection*{An example with restricted strategic spaces}

New outcomes are possible when quantum strategies are confronted. Consider a restricted strategic space in which  $\protect{p_R+p_S=1}$, so a players strategy is (aside from quantum phases) determined by a single angular parameter $\xi$ defined by $\cos^2\xi\equiv p_R$. For $\xi=0$ $(p_R=1)$ it reduces to Pavlov and for $\xi=\pi/4$, $\protect{(p_R=0.5)}$ to Random. Values of $\xi\in[0,\pi/4]$ result in strategies which interpolate between Random and Pavlov. If the same parametrization is adopted for Bob's strategy, the resulting two-parameter coin operation (assuming Alice plays first) is $\protect{U_c=U_B(\xi_B)\cdot U_A(\xi_A)}$.
%
\begin{figure}
\psfrag{alpha}[][][1.5]{\vspace{2mm}\Huge$\xi_A$}
\psfrag{beta}[][][1.5]{\hspace{-1mm}\Huge$\xi_B$}
\psfrag{Payoff}[][][1.5]{\Huge$\bar x_{A,B}$}
\psfrag{x0}[][bl][1.3]{\Huge$~0$}\psfrag{xps4}[][bl][1.3]{\Huge$~\frac{\pi}{4}$}
\psfrag{y0}[][bl][1.3]{\Huge$0\quad$}\psfrag{yps4}[][bl][1.3]{\Huge$\frac{\pi}{4}\quad$}
   \begin{center}
\resizebox{6.6cm}{!}{\includegraphics[angle=-90]{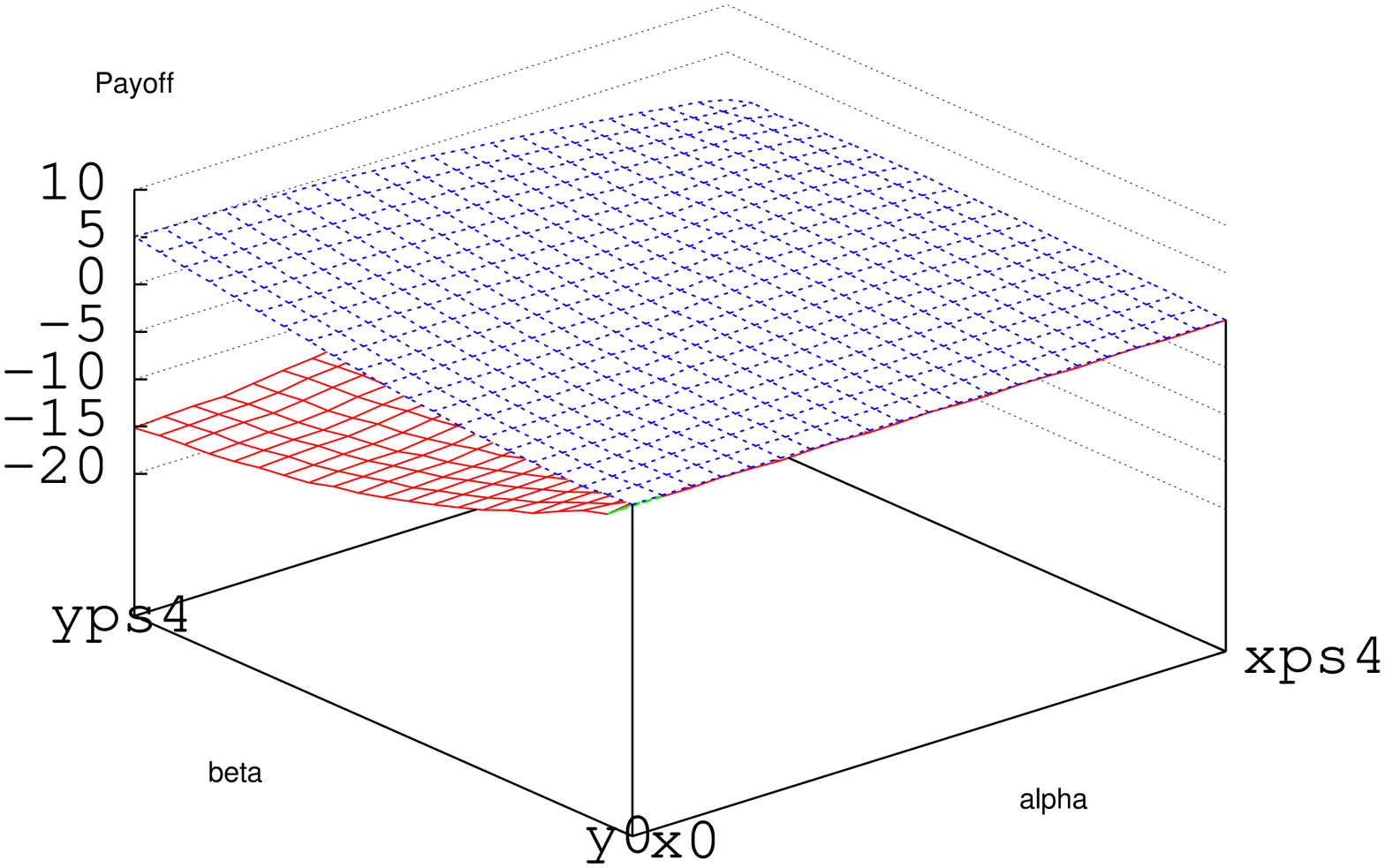}}\hspace{5mm}
\resizebox{6.6cm}{!}{\includegraphics[angle=-90]{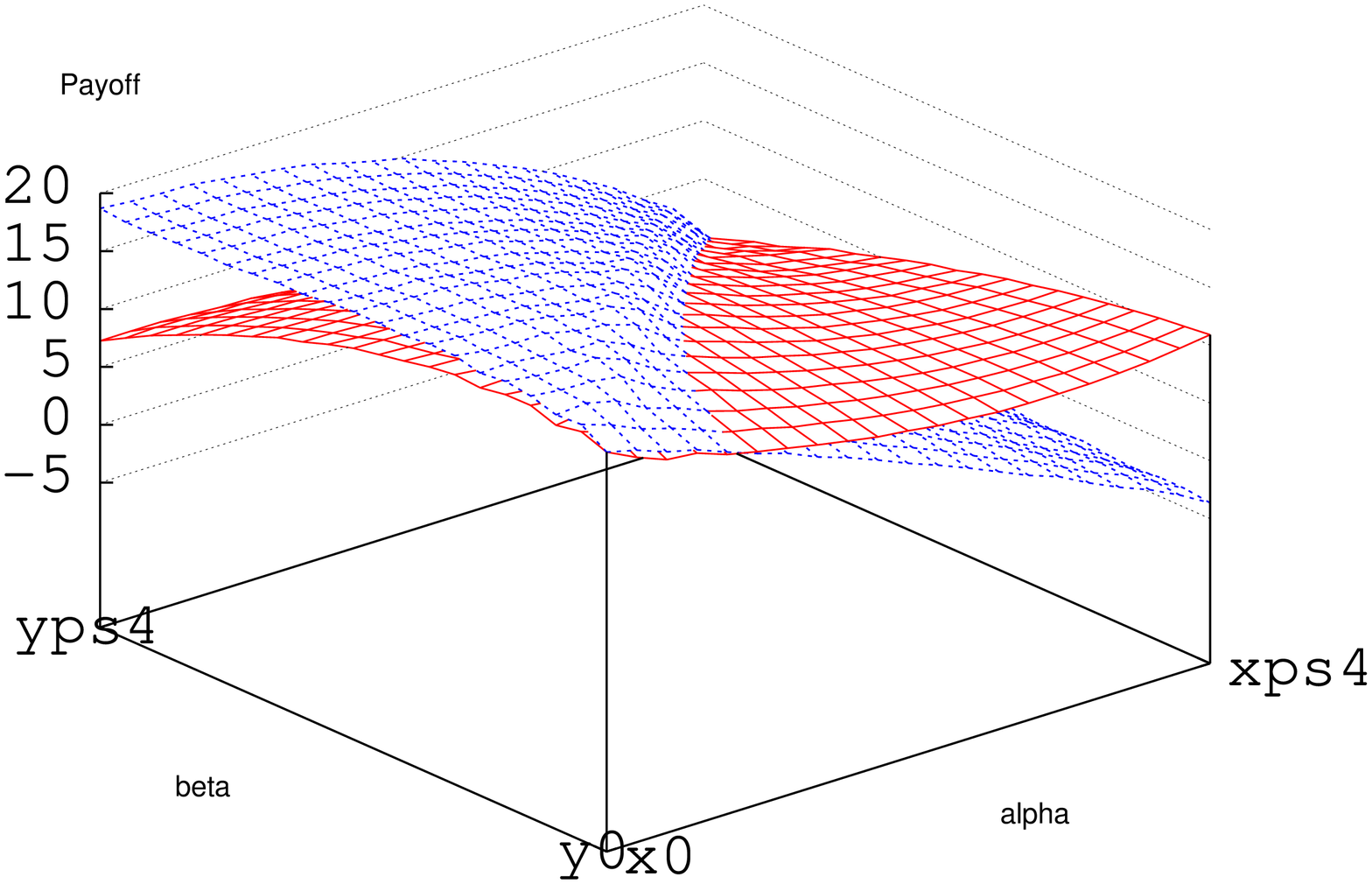}}
   \end{center}
\caption{\footnotesize  (color online) Alice's (red) and Bob's (blue) payoffs after $N=50$ steps as a function of the strategic choice parametrized by angles $\xi_{A,B}\in[0,\pi/4]$ defined by $\cos^2\xi\equiv p_R$ for each player. The surfaces correspond to two unbiased initial conditions (see text for details). }
\label{fig:alice-param-surf}
\end{figure}
This surface, after $N=50$ iterations, is shown in Fig.~\ref{fig:alice-param-surf} for two unbiased initial coin states: the product state $\protect{(\ket{00}+i\ket{01}+i\ket{10}-\ket{11})/2}$ (left panel)  and the fully entangled state $(\ket{00}+\ket{11}$)/$\sqrt{2}$ (right panel). Our results are for a set of unbiased values of the parameters which fulfill the PD constrains, eq.~(\ref{ineq}), $R$=$-P$=1 and $T$=$-S$=2. These results show that the classical situation (a tie for unbiased initial conditions) is exceptional in the quantum case and in the quantum game different initial coin states result in very different outcomes. 

\begin{figure}[ht]
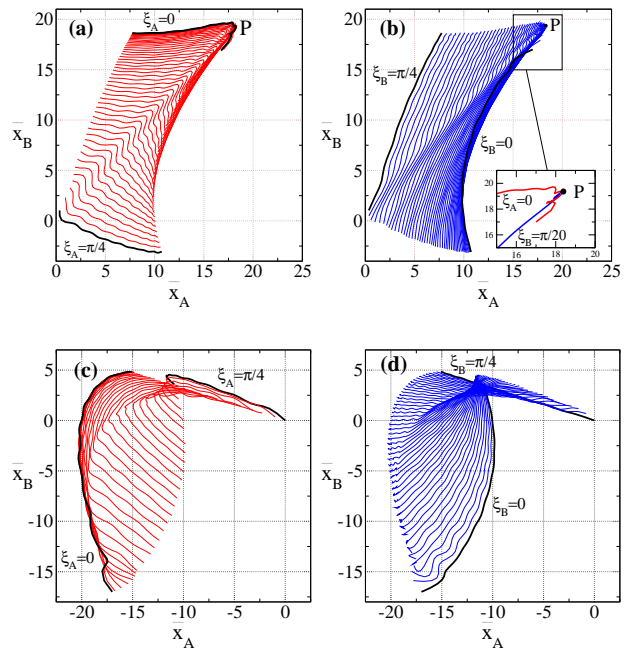

   \begin{center}
     \includegraphics[height=4cm,width=4cm]{./fig/ia-bell00.eps}~\includegraphics[height=4cm,width=4cm]{./fig/ib-bell00.eps}\\\vskip5mm
 \includegraphics[height=4cm,width=4cm]{./fig/ia-bell01.eps}~\includegraphics[height=4cm,width=4cm]{./fig/ib-bell01.eps}
   \end{center}
\caption{Average cumulated payoffs $(\bar x_A \mbox{ vs. } \bar
x_B)$ after $50$
iterations for different initial conditions. In panels (a,b) the
initial coin is $\protect{\left(\ket{00}+\ket{11}\right)/\sqrt{2}}$ and in
panels (c,d)    $\protect{\left(\ket{01}+\ket{10}\right)/\sqrt{2}}$
Panels (a, c) show lines of constant $\xi_A$ (red online) and
panels (b,d) show lines of constant $\xi_B$ (blue online). The
extreme values of $\xi_A$ and $\xi_B$ are indicated by thick lines.
In both cases, the other parameter ranges from $0$ (a Pavlovian
strategy) to $\pi/4$ (a random strategy). The inset in panel (b)
shows the lines for $\xi_A=0, \xi_B=\pi/20$ which intersect at the
point P in which both agents maximize their average payoffs.}
\label{fig:XvsY}
\end{figure}

It is illustrative to look at the payoff of one agent vs the payoff
of the other $(\bar x_A$ vs. $\bar x_B)$. In Fig.~\ref{fig:XvsY} we
show the results for two different unbiased initial coin states,
$\protect{\left(\ket{00}+\ket{11}\right)/\sqrt{2}}$ and
$\protect{\left(\ket{01}+\ket{10}\right)/\sqrt{2}}$

Note that, independently of Bob's  choice of strategy $\xi_B$, Alice
must play Pavlov ($\xi_A=0$) in order to maximize her average
return (see panel (a) ). Bob gets the highest average payoff when he
adopts an intermediate strategy with $\protect{\xi_B\simeq\pi/20}$, provided
Alice plays Pavlov (see panel (b) ). In fact, this point
$\protect{(\xi_A,\xi_B)=(0,\pi/20)}$ is a Nash equilibrium which is also
Pareto optimal (point P in the inset of Fig.~\ref{fig:XvsY}b). As
shown in panels (c,d), the same Pavlov strategy may result in a maximum or minimum payoff for Alice, depending on
Bob's choice of strategy.

As discussed above, not all classical strategies have quantum analogs in a sequential iterated quantum game. Clearly, playing always C (or D) is forbidden because it leads to non unitary operations. For the same reason, the TFT strategy is also forbidden in a sequential quantum game. 
However, if the strategic space is be extended by considering {\it simultaneous} moves of both players, this strategy becomes an option. 

\subsection{Simultaneous strategies}

The case of simultaneous moves is closer to the classical situation and allows some new strategies to be implemented. Let $[p_R^A,p_S^A,p_T^A,p_P^A]$ define the classical strategy to be implemented by Alice and $[p_R^B,p_S^B,p_T^B,p_P^B]$ the one by Bob. A simultaneous quantum game confronting quantum versions of two these classical strategies involves unitary  operations of the form, 
$$
\left[
\begin{array}{cccc}
e^{i\varphi_{11}}\sqrt{p_R^Ap_R^B} &e^{i\varphi_{12}}\sqrt{p_R^A\bar p_R^B}  &e^{i\varphi_{13}} \sqrt{\bar p_R^A p_R^B} &e^{i\varphi_{14}} \sqrt{\bar p_R^A\bar p_R^B} \\
e^{i\varphi_{21}}\sqrt{p_S^Ap_T^B} &e^{i\varphi_{22}}\sqrt{p_S^A\bar p_T^B}  &e^{i\varphi_{23}} \sqrt{\bar p_S^A p_T^B} & e^{i\varphi_{24}}\sqrt{\bar p_S^A\bar p_T^B}\\
e^{i\varphi_{31}}\sqrt{p_T^Ap_S^B} &e^{i\varphi_{32}}\sqrt{p_T^A\bar p_S^B}  & e^{i\varphi_{33}}\sqrt{\bar p_T^A p_S^B} & e^{i\varphi_{34}}\sqrt{\bar p_T^A\bar p_S^B} \\
e^{i\varphi_{41}}\sqrt{p_P^Ap_P^B} &e^{i\varphi_{42}}\sqrt{p_P^A\bar p_P^B}  & e^{i\varphi_{43}}\sqrt{\bar p_P^A p_P^B} & e^{i\varphi_{44}}\sqrt{\bar p_P^A\bar p_P^B}
\end{array}
\right]
$$
where $e^{i\varphi_{kl}}$ are phase factors and $\bar p_i^{A,B}\equiv 1-p_i^{A,B}$. For a real $U_c$, $e^{i\varphi_{kl}}=\pm 1$, unitarity requires 
\begin{eqnarray*}
(p_R^A-\bar p_S^A)(p_R^B-\bar p_T^B)&=&(p_R^A-\bar p_T^A)(p_R^B-\bar p_S^B)=0,\nonumber\\
(p_R^A-\bar p_P^A)(p_R^B-\bar p_P^B)&=&(p_S^A-\bar p_T^A)(p_S^B-\bar p_T^B)=0,\label{simcond}\\
(p_S^A-\bar p_P^A)(p_T^B-\bar p_P^B)&=&(p_T^A-\bar p_P^A)(p_S^B-\bar p_P^B)=0.\nonumber
\end{eqnarray*}
In the general case, analogous restrictions involving the phases $\varphi_{kl}$ apply. For instance, the game in which Alice plays Pavlov $[1,0,0,1]$ and Bob simultaneously plays TFT $[1,0,1,0]$ is implemented by 
$$
U_c^{PT} = \opp{00}{00}+ e^{i\lambda_1}\!\opp{10}{01}+e^{i\lambda_2}
\opp{11}{10}+e^{i\lambda_3}\opp{01}{11},\label{UBTFT} 
$$
in terms of three arbitrary phases $\lambda_i$. In the classical version of this game, if both agents start playing C with probability $0.5$, after $N$ iterations each collects a null average payoff $N(R+P)=0$. In the quantum game with the above mentioned Bell state as initial condition, the winning chances are not equal and both players end up with positive payoffs. In a similar way, other classical strategies may be confronted. The simultaneous scheme is closer to classical games but it is limited by conditions (\ref{simcond}). For example, if both players adopt {\it Pavlov}, they cannot play simultaneously, as some of these conditions are not satisfied. In Table~I, we consider three classical strategies and indicate which of them can be confronted within the quantum sequential and/or simultaneous schemes. Games confronting TFT vs Random may be described by (non-unitary) quantum operations. We do not consider these extensions in this work. 

\begin{table}
\begin{tabular}{|l|c|c|c|}
\hline
 &Random & Pavlov & TFT \\\hline
Random & 1, 2 & 1,2 & not unitary\\
Pavlov & 1,2 & 1 & 2 \\
TFT & not unitary & 2 & 2 \\\hline
\end{tabular}
\caption{\small Some strategies that can be confronted both in sequential (1) and simultaneous (2) quantum games.}
\end{table}

\section{Concluding Remarks} 

We have related general bi-partite iterated quantum games to  discrete time quantum walks. Several strategies from classical game theory can be implemented in terms of elementary two-qubit quantum gates. Each of them gives rise to a family of quantum strategies. We give the conditions that must be satisfied so that two classical strategies may be confronted, either sequentially or simultaneously, in an interated quantum game. Some well-known classical strategies, such as TFT can only be implemented in the simultaneous scheme. Non-conmuting operations, such as those associated to a Pavlov-Pavlov confrontation, can only be realized in the sequential scheme. Since the parameter space for these quantum strategies is extremely large, instead of a systematic exploration, we have shown through selected examples that the outcome of a QG may be different from that of the classical counterpart. 

In one-shot quantum games, there is a threshold for the amount of entanglement in the initial state that allows quantum features to emerge \cite{Eisert99,Du03}. In our proposal, entanglement is dynamically generated by conditional operations and the preparation of an initially entangled state is not required. We have characterized the bi-partite entanglement between both agents in a Pavlov-Random QG using the von Neumann entropy of the reduced density operator (entropy of entanglement) and found that this quantity increases at a logarithmic rate. In order to exploit entanglement partial measurements may be included as part of the strategic choices.

The connection between bipartite quantum games and discrete-time quantum walks opens the possibility of experimentally
testing iterated quantum games and strategies using simple linear optics elements \cite{Do}. The sensitivity of these QGs to the choice of the initial state may be attenuated in experimental realizations through the introduction of decoherence. The impact of a weak coupling to the environment is a relevant issue in the study of quantum games,  which
deserves further study, as, in the classical case, noise-related effects are able to radically change the outcome of the different strategies \cite{Nowak}. An initial step in this direction may be considering the outcome, for different strategic options, of opposing a classical player vs. a quantum player. 

The scheme we have introduced for quantizing the iterated PD game can obviously be applied to  $2\times 2$ games with an arbitrary payoff matrix. There are several popular games that seem interesting to analyze within this framework. For example the {\it Hawk-Dove} \cite{ms82}, in which the damage from mutual defection in the PD is increased so that it finally exceeds the damage suffered by being exploited: $T>R>S>P$. Or the {\it Stag Hunt} game \cite{s04}, corresponding to the payoffs rank order $R>T>P>S$ {\it i.e.} when the reward $R$ for mutual
cooperation in the PD games surpasses the temptation $T$ to defect.

Another generalization of this scheme involves multi-partite games. The basic evolution, given by eqs.~(\ref{evol}) and (\ref{shift-op-gen}), may be generalized to accomodate any number of quantum walkers. This may be useful for the ''public goods`` problem, since there are indications that the quantum version of this multiplayer game may provide a more efficient distribution of resources \cite{ChenHogg06}. However, this generalization raises non-trivial issues regarding the multipartite entanglement which may be dynamically generated within the game. 

\section*{Aknowledgements}
\noindent
{\it Work supported by PEDECIBA and PDT project 29/84 (Uruguay), CNPq and FAPERJ (Brazil).}

\end{document}